\documentclass[8.5pt,twoside,twocolumn]{article}
\usepackage[super,sort&compress,comma]{natbib} 
\usepackage{mhchem}
\usepackage{times,mathptmx,amssymb}
\usepackage{sectsty}
\usepackage{balance} 

\usepackage{graphicx} 
\usepackage{lastpage}
\usepackage[format=plain,justification=raggedright,singlelinecheck=false,font=small,labelfont=bf,labelsep=space]{caption} 
\usepackage{fancyhdr}
\usepackage{color}
\begin{document}

\thispagestyle{plain}

\makeatletter 
\def\subsubsection{\@startsection{subsubsection}{3}{10pt}{-1.25ex plus -1ex minus -.1ex}{0ex plus 0ex}{\normalsize\bf}} 
\def\paragraph{\@startsection{paragraph}{4}{10pt}{-1.25ex plus -1ex minus -.1ex}{0ex plus 0ex}{\normalsize\textit}} 
\renewcommand\@biblabel[1]{#1}            
\renewcommand\@makefntext[1]%
{\noindent\makebox[0pt][r]{\@thefnmark\,}#1}
\makeatother 
\renewcommand{\figurename}{\small{Fig.}~}
\sectionfont{\large}
\subsectionfont{\normalsize} 

\fancyfoot{}
\fancyfoot[LO,RE]{\vspace{-7pt}\includegraphics[height=9pt]{headers/LF}}
\fancyfoot[CO]{\vspace{-7.2pt}\hspace{12.2cm}\includegraphics{headers/RF}}
\fancyfoot[CE]{\vspace{-7.5pt}\hspace{-13.5cm}\includegraphics{headers/RF}}
\fancyfoot[RO]{\footnotesize{\sffamily{1--\pageref{LastPage} ~\textbar  \hspace{2pt}\thepage}}}
\fancyfoot[LE]{\footnotesize{\sffamily{\thepage~\textbar\hspace{3.45cm} 1--\pageref{LastPage}}}}
\fancyhead{}
\renewcommand{\headrulewidth}{1pt} 
\renewcommand{\footrulewidth}{1pt}
\setlength{\arrayrulewidth}{1pt}
\setlength{\columnsep}{6.5mm}
\setlength\bibsep{1pt}

\twocolumn[
  \begin{@twocolumnfalse}
\noindent\LARGE{\textbf{Cooperative wrapping of nanoparticles by membrane tubes}}
\vspace{0.6cm}

\noindent\large{\textbf{Michael Raatz, Reinhard Lipowsky, and Thomas R.\ Weikl}}


\vspace{0.6cm}

\noindent \normalsize{
The bioactivity of nanoparticles crucially depends on their ability to cross biomembranes. Recent simulations indicate the cooperative wrapping and internalization of spherical nanoparticles in tubular membrane structures. In this article, we systematically investigate the energy gain of this cooperative wrapping by minimizing the energies of the rotationally symmetric shapes of the membrane tubes and of membrane segments wrapping single particles. We find that the energy gain for the cooperative wrapping of nanoparticles in membrane tubes  relative to their individual wrapping as single particles strongly depends on the ratio $\rho/R$ of the particle radius $R$ and the range $\rho$ of the particle-membrane adhesion potential. For a potential range of the order of one nanometer, the cooperative wrapping in tubes is highly favorable for particles with a radius of tens of nanometers and intermediate adhesion energies,  but not for particles that are significantly larger.
}
\vspace{0.5cm}
 \end{@twocolumnfalse}
  ]

\footnotetext{\textit{Max Planck Institute of Colloids and Interfaces, Department of Theory and Bio-Systems, Science Park Golm, 14424 Potsdam, Germany.}}

\section{Introduction}

Advances in nanotechnology have led to an increasing interest in how nanoparticles interact with living organisms \cite{Nel09,DeJong08}. To enter the cells or cell organelles of such organisms, nanoparticles have to cross biomembranes. This crossing or internalization requires (i) the wrapping of the particles by the membrane and (ii) the subsequent fission of a membrane neck if the particles are larger than the membrane thickness and cannot cross the membrane directly. In general, both wrapping and fission can either be passive \cite{Rothen06,Liu11}, or can be actively driven or assisted by protein machineries that consume chemical energy \cite{Mukherjee97,Conner03,Hurley10,Rodriguez13}.  Passive wrapping can occur if the adhesive interaction between the nanoparticles and membranes is sufficiently strong to compensate for the cost of membrane bending. The passive wrapping of nanoparticles has been investigated in experiments with lipid vesicles\cite{Dietrich97,Koltover99,Fery03,LeBihan09,Michel12}, polymersomes \cite{Jaskiewicz12,Jaskiewicz12b}, and cells \cite{Rothen06,Liu11}.

\begin{figure*}[tp]
\centering
\includegraphics[width=1.8\columnwidth]{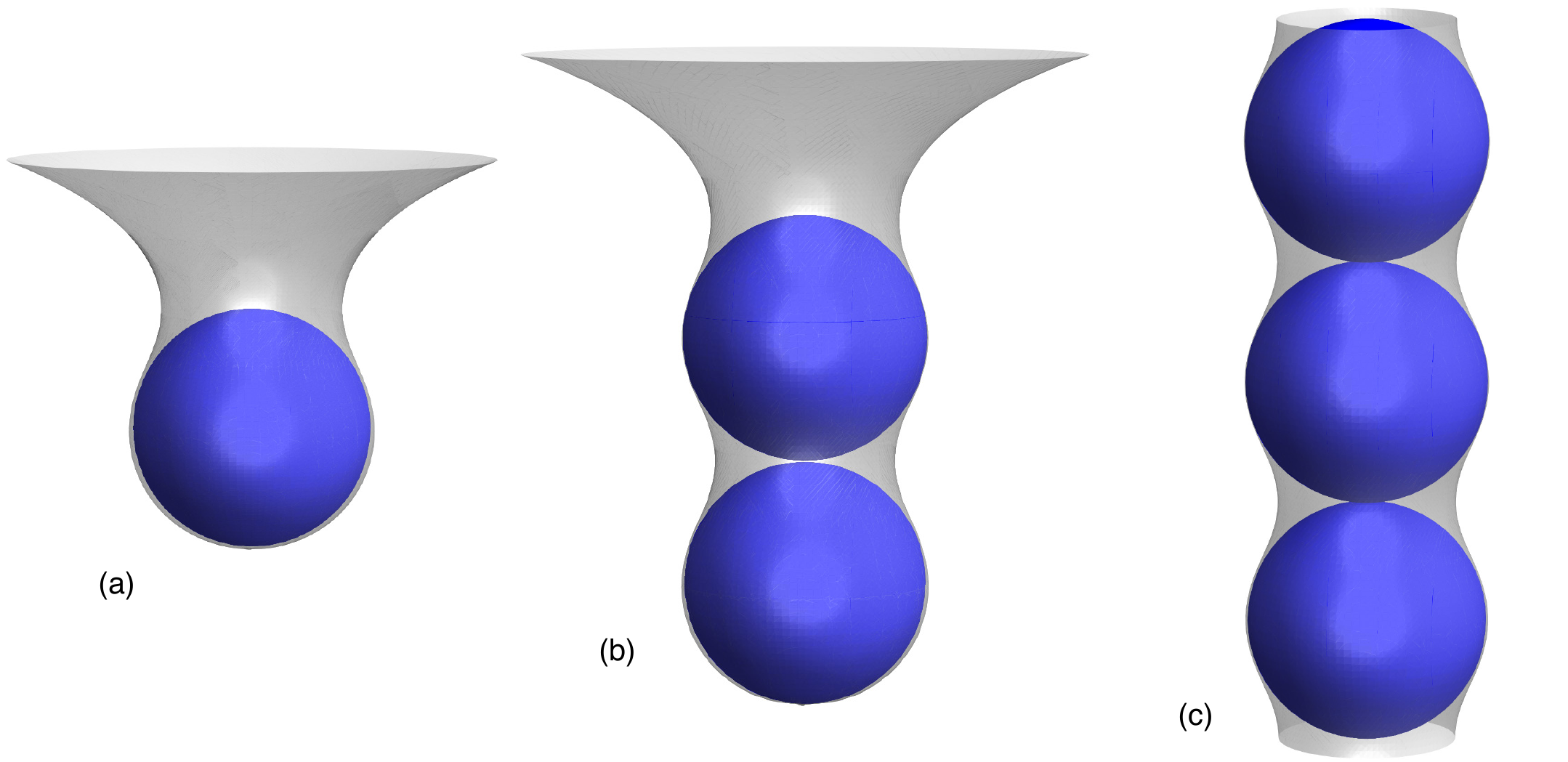}
\caption{Minimum-energy states of (a) a single spherical particle wrapped by a tensionless membrane, (b) two particles wrapped in a membrane tube, (c) three central particles of a long membrane tube for the range $\rho = 0.1 R$ of the particle-membrane adhesion potential and the rescaled adhesion energy $u= U R^2/\kappa$ = 3 where $R$ is the particle radius and $\kappa$ is the bending rigidity of the membrane.  The minimum total energies of these states are (a) $E = -15.85 \,\kappa$, (b) $E = -17.19 \,\kappa$ per particle, and (c) $E = -18.53 \,\kappa$ per particle. The total energy is the sum of the adhesion energy of the particles and the bending energy of the membrane. 
}
\label{figure_3Dshapes}
\end{figure*}

While theoretical\cite{Lipowsky98,Boulbitch02,Deserno03,Deserno04,Fleck04,Gozdz07,Benoit07,Decuzzi08,Chen09,Yi11,Cao11,Dasgupta13,Bahrami13}  and simulation\cite{Noguchi02,Smith07,Fosnaric09,Li10,Yang10,Vacha11,Shi11} efforts have been largely focused on the passive wrapping of single nanoparticles, recent simulations indicate the cooperative wrapping of several nanoparticles in tubular membrane structures \cite{Bahrami12,Saric12b,Yue12}. To better understand the formation of these particle-filled membrane tubes, we investigate here the energy gain for the cooperative wrapping of nanoparticles in tubes by minimizing the energies of the rotationally symmetric shapes of membrane tubes and of membrane segments wrapping single particles. We find that this energy gain strongly depends on the ratio $\rho/R$ of the particle radius $R$ and the range $\rho$ of the particle-membrane adhesion potential. As examples, Fig.~\ref{figure_3Dshapes} displays minimum-energy shapes of a tensionless membrane for the potential range $\rho = 0.1 R$ and the rescaled adhesion energy $u= U R^2/\kappa = 3$ where $R$ is the particle radius and $\kappa$ is the bending rigidity of the membrane. The minimum energy is $E = -15.8 \,\kappa$ for the single wrapped particle in  Fig.~\ref{figure_3Dshapes}(a),  $E = -17.2 \,\kappa$ per particle for the membrane tube of two particles in Fig.~\ref{figure_3Dshapes}(b), and $E = -18.5 \,\kappa$ per particle for the central particles of a long membrane tube in Fig.~\ref{figure_3Dshapes}(c). The energy gain for the cooperative wrapping of particles in a long tube, compared to the individual wrapping of these particles, thus is $\Delta E = -2.7 \,\kappa$ per particle. At the same rescaled adhesion energy $u=3$, this energy gain per particle reduces to $\Delta E = -0.76 \,\kappa$ for the potential range  $\rho = 0.03 R$, to $\Delta E = -0.24 \,\kappa$ for $\rho = 0.01 R$, and vanishes as $\rho/R$ approaches zero. For a potential range of the order of one nanometer and typical bending rigidities of lipid membranes between 10 and 20 $k_B T$ \cite{Seifert95}, the cooperative wrapping in membrane tubes thus is highly favorable for nanoparticles with a radius in the range of tens of nanometers since the energy gain $\Delta E$ then is significantly larger than the thermal energy $k_B T$, provided these particles do not strongly repel each other. 

This article is organized as follows: In section 2, we introduce our model and our minimization method, which is  based on a discretization of the profiles of the rotationally symmetric membrane shapes shown in Fig.~\ref{figure_3Dshapes}. In section 3, we consider the wrapping of a single particle by a tensionless membrane, and describe how the wrapping degree of the particle and the minimum energy of the membrane depend on the ratio $\rho/R$ of the potential range $\rho$  and particle radius $R$ and on the rescaled adhesion energy $u=U R^2/\kappa$, which characterizes the relative strength of adhesion and bending. In section 4, we investigate the cooperative wrapping of particles in long membrane tubes and determine the energy gain $\Delta E$ per particle relative to individual wrapping as a function of $u$ and $\rho/R$. In section 5, we consider the tubular membrane structures induced by two or more particles. The article ends with a discussion and conclusions.

\section{Model and minimization method}

\begin{figure}[tp]
\centering
\includegraphics[width=0.9\columnwidth]{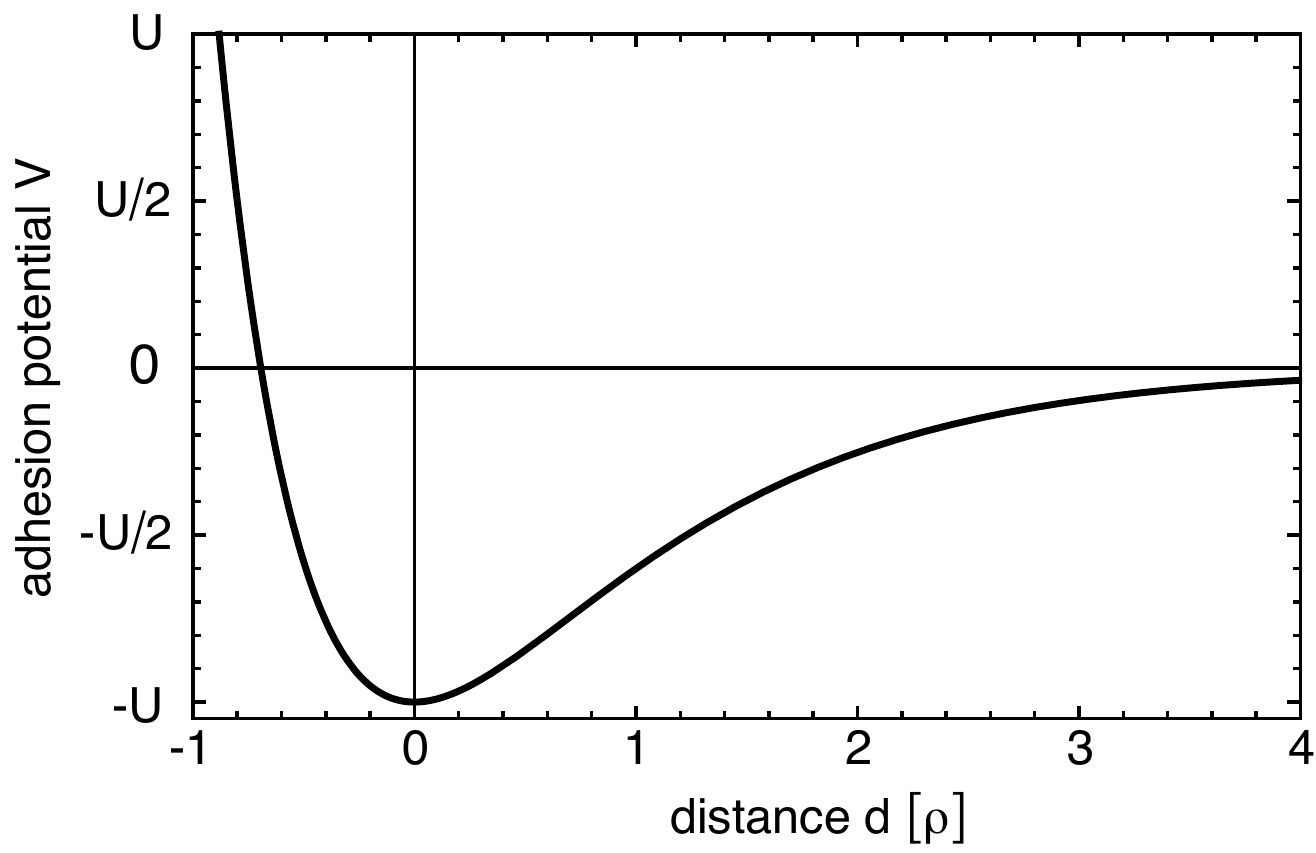}
\caption{Adhesion potential $V$ as a function of the relative distance $d$ of the membrane from the particle surface (see Eq.~(\ref{V})). The adhesion potential has a minimum of depth $U$ at the relative distance $d=0$,which corresponds to the equilibrium distance of a bound membrane patch. The range of the adhesion potential is denoted by $\rho$.
 }
\label{figure_potential}
\end{figure}

The passive wrapping of particles by a membrane is governed by the interplay of bending and adhesion \cite{Lipowsky98,Seifert90}. The total energy $E$ is the sum 
\begin{equation}
E = E_\text{be} + E_\text{ad}
\end{equation}
of the bending energy $E_\text{be}$ of the membrane and the adhesion energy $E_\text{ad}$ of the particles. 
The bending energy of the membrane is the integral
\begin{equation}
E_\text{be} = 2\kappa \int M^2\, \rm{d}A
\label{Ebe}
\end{equation}
over the area $A$ of the membrane with local mean curvature  $M$ and bending rigidity $\kappa$ \cite{Helfrich73}. We assume here that the membrane has a spontaneous curvature of zero, and neglect a constant term in the bending energy from the integral of the Gaussian curvature. The adhesion energy of the membrane in contact with $n_p$ particles is the integral
\begin{equation}
E_\text{ad} = \int \sum_{i=1}^{n_p} V(d_i) dA 
\label{Ead}
\end{equation}
with an adhesion potential $V$ that depends on the local relative distance $d_i$ of the membrane from particle $i$. The adhesion potential considered in this article has the functional form
\begin{equation}
V(d_i) = U \left(e^{-2d_i/\rho} - 2 \,e^{-d_i/\rho}\right)
\label{V}
\end{equation}
of a Morse potential with characteristic potential depth $U$ and range $\rho$. The potential $V(d_i)$ adopts its minimum value $-U$ at the relative distance $d_i=0$  (see Fig.~\ref{figure_potential}).  The relative distance $d_i=0$ thus corresponds to the equilibrium distance between a particle and a bound membrane patch in the absence of other than adhesive forces.

The four parameters of our model are the bending rigidity $\kappa$ of the membrane, the potential depth $U$ and range $\rho$ of the particle-membrane interaction (\ref{V}), and the radius $R$ at which the adhesion energy of membrane segments bound to the spherical particles is minimal. If the bound membrane is in direct contact with the particle, this radius is approximately $R \simeq R_p + d_m/2$ where $R_p$ is the actual particle radius and $d_m$ is the membrane thickness, because our membrane profiles correspond to the membrane midplanes. For simplicity, the radius $R$ here is denoted as particle radius. Since we are free to choose both an energy scale and a length scale as units of energy and length in our model, the four parameters $\kappa$, $U$, $\rho$, and $R$ of our model can be reduced to two independent, dimensionless parameters. We choose here as independent parameters the ratio $\rho/R$ of the potential range $\rho$ and radius $R$ and the rescaled adhesion energy $u = U R^2/\kappa$, which characterizes the relative strength of adhesion and bending.

\thispagestyle{plain}

Our aim here is to determine the minimum-energy shapes of the membranes around a single spherical particle and around linear aggregates of particles in tubular structures. Since these membrane shapes are rotationally symmetric, we describe the membranes as surfaces of revolution using two different parametrizations.

In {\em parametrization 1}, the rotationally symmetric membrane shapes are described by the local radial distance $r$ as a function of the coordinate $z$ along the axis of rotation:
\begin{equation}
\vec{r}(z,\phi) = \left(
\begin{array}{c}
r(z)\cos\phi\\
r(z)\sin\phi\\
z
\end{array}
\right)
\end{equation}
Here, $\vec{r}(z,\phi)$ is the vector of cartesian coordinates for a point on the membrane surface with $0\le \phi < 2 \pi$. We use this parametrization to describe e.g.\ the rotationally symmetric shapes around central particles in long tubular membrane structures (see Fig.~\ref{figure_3Dshapes}(c)). In this parametrization, the bending energy (\ref{Ebe}) and adhesion energy (\ref{Ead}) adopt the form 
\begin{equation}
E_\text{be}=\pi \kappa\int \frac{\left(r(z)r^{\prime\prime}(z)-r^\prime(z)^2-1\right)^2}{r(z)\left(r^\prime(z)^2+1\right)^{5/2}}\text{d}z
\label{Ebe_1}
\end{equation}
\begin{equation}
E_\text{ad} = 2\pi\int  \sum_{i=1}^{n_p} V(d_i)r(z)\sqrt{1+r^\prime(z)^2}\text{d}z
\label{Ead_1}
\end{equation}
with $d_i = d_i(z,r(z))$. The primes here indicate derivatives with respect to $z$.

\thispagestyle{plain}

\begin{figure}[tp]
\centering
\includegraphics[width=0.98\columnwidth]{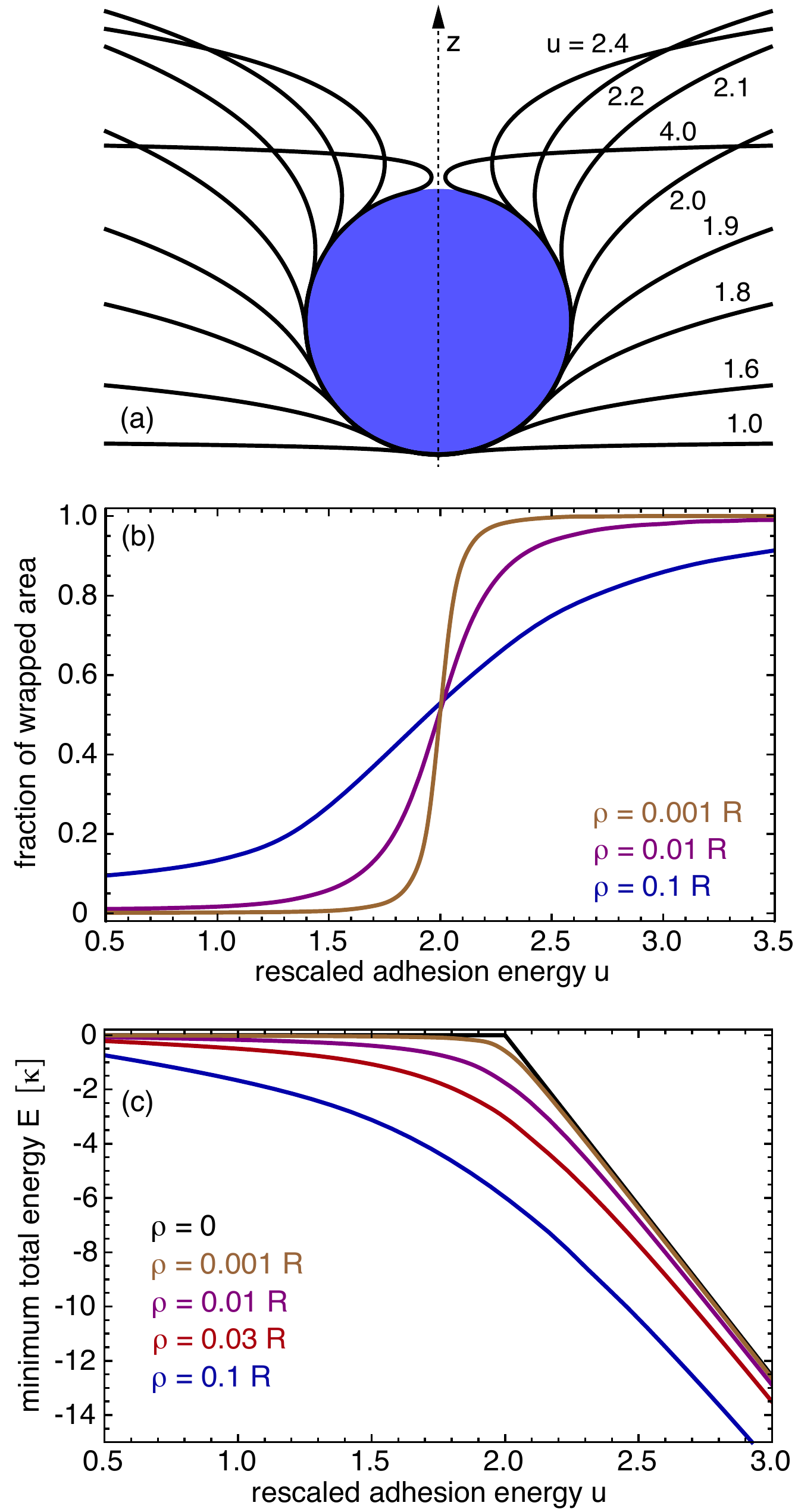}
\caption{(a) Minimum-energy profiles of the rotationally symmetric membrane shapes around a single spherical particle for the potential range $\rho = 0.01 R$. The numbers indicate the values for the rescaled adhesion energy $u$ of the different profiles. The shapes are axially symmetric with respect to the $z$-axis (dashed arrow). (b) Area fraction of a single spherical particle that is wrapped by the membrane as a function of the rescaled adhesion energy $u$ for three different values of the potential range $\rho$. The wrapped area fraction is determined from a projection of those membrane segments that have a distance smaller than $R+\rho$ from the particle center on a sphere with radius $R$. The distance $R$ from the particle center corresponds to the minimum of the adhesion potential. 
(c) Minimum total energy $E$ of the membrane around a single particle as a function of the rescaled adhesion energy $u$ for different values of the potential range $\rho$.
}
\label{figure_one}
\end{figure}

In {\em parametrization 2}, we describe the rotationally symmetric shapes by the height $z$ along the axis of rotation as a function of the radial distance $r$ from this axis:
\begin{equation}
\vec{r}(r,\phi) = \left(
\begin{array}{c}
r\cos\phi\\
r\sin\phi\\
z(r)
\end{array}
\right)
\end{equation}
We use this parametrization to describe partially wrapped states of a single particle in which up to half of the particle surface is wrapped by the membrane. In this parametrization, the bending energy and adhesion energy adopt the form
\begin{eqnarray}
E_\text{be} &=& \pi\kappa \int\frac{(r z^{\prime\prime}(r) + z^\prime(r)^3+z^\prime(r))^2}{r(1+z^\prime(r)^2)^{5/2}}\text{d}r\\
E_\text{ad} &=& 2\pi\int  \sum_{i=1}^{n_p} V(d_i)r\sqrt{1+z^\prime(r)^2}\,\text{d}r
\end{eqnarray}

\thispagestyle{plain}

In addition, we use a combination of both parametrizations to describe deeply wrapped states of a single particle and the wrapping of two particles by a membrane tubule. In this combination, the membrane is divided into two parts that are described by parametrization 1 and 2, respectively. 

For a numerical minimization of the total  energy $E=E_\text{be}+E_\text{ad}$, we discretize the functions $r(z)$ and $z(r)$ of the two parametrizations using up to 1000 discretization points and express the first and second derivatives of these functions as finite differences. We obtain the minimum-energy shapes then from constrained minimization with respect to the functional values at the discretization points using the program Mathematica \cite{Mathematica8}.

\thispagestyle{plain}

\section{Wrapping of a single particle}

We first consider the wrapping of a single spherical particle by a tensionless membrane. Fig.~\ref{figure_one}(a) displays the minimum-energy membrane profiles around a single particle for different values of the rescaled adhesion energy $u$ between $1.0$ and $4.0$  and for the potential range $\rho = 0.01 \, R$. The profiles consist of a bound membrane segment with circular profile that is wrapped around the particle and an unbound membrane segment with a profile that eventually approaches the planar membrane, which is oriented perpendicular to the rotational symmetry axis of the membrane shapes. We assume that the planar membrane  is large and, thus, constitutes an area reservoir for wrapping. 

In Fig.~\ref{figure_one}(b), the fraction of the particle's surface area that is wrapped by the membrane is displayed as a function of the rescaled adhesion energy $u$ for three different potential ranges. The fraction of of the wrapped particle area  continuously increases with $u$. The continuous wrapping process is centered around the value $u=2$ of the rescaled adhesion energy. At this value of $u$, the adhesion energy $E_\text{ad} = - 4 \pi  R^2 U\, x$ of a spherical membrane segment that is located in the minimum of the adhesion potential and wraps the fraction $x$ of the particle surface is equal to the bending energy $E_\text{be} = 8 \pi \kappa\, x$ of this segment \cite{Lipowsky98}. With decreasing potential range $\rho$, the wrapping process becomes more abrupt (see Fig.~\ref{figure_one}(b)).

\begin{figure}[tp]
\centering
\includegraphics[width=\columnwidth]{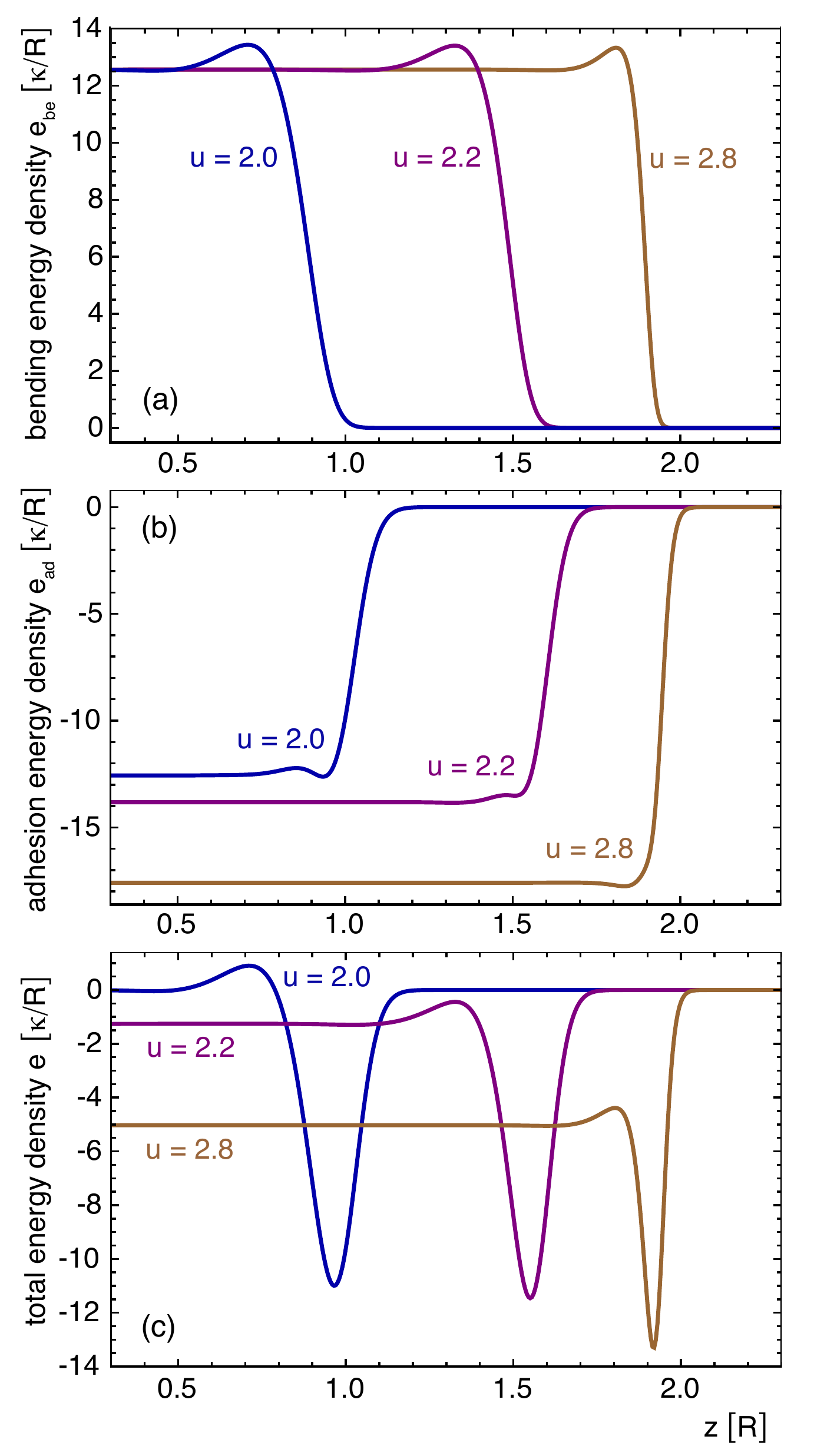}
\caption{(a) Bending energy density $e_\text{be}(z)$, (b) adhesion energy density $e_\text{ad}(z)$, and (c) total energy density $e(z) = e_\text{be}(z)+ e_\text{ad}(z)$  for the two shape profiles in Fig.~\ref{figure_one}(a) with rescaled adhesion energies $u = 2.0$ and 2.2 and a third shape profile with $u = 2.8$. Here, $z$ is the coordinate along the axis of rotation indicated in Fig.~\ref{figure_one}(a). The value $z=R$ corresponds to the center of the spherical particle.  The bending and adhesion energies are related to the energy densities via the integrations $E_\text{be} = \int  e_\text{be}(z)\text{d}z$ and  $E_\text{ad} = \int  e_\text{ad}(z)\text{d}z$. 
}
\label{figure_one_energy-profiles}
\end{figure}

\thispagestyle{plain}

Fig.~\ref{figure_one_energy-profiles} displays the energy densities along the rotational symmetry axis for some of the shape profiles of Fig.~\ref{figure_one}(a). The center of the particle is located at the value $z = R$ of the coordinate along the symmetry axis. The bending energy density is the integrand $e_\text{be}(z)=\text{d}E_\text{be}/\text{d}z$ of the bending energy (\ref{Ebe_1}) obtained in parametrization 1. In this parametrization, the membrane profile is described by the function $r(z)$ where $r$ is the radial distance of the membrane from the axis of rotation. The adhesion energy density is the integrand $e_\text{ad}(z)=\text{d}E_\text{ad}/\text{d}z$ of the adhesion energy (\ref{Ead_1}), and the total energy density is the sum $e(z) = e_\text{be}(z)+e_\text{ad}(z)$ of the bending and adhesion energy densities. The bending and adhesion energy densities of Fig.~\ref{figure_one_energy-profiles} initially adopt the constant values $e_\text{be} = 4 \pi \kappa /R$ and $e_\text{ad} = -2 \pi u\kappa/ R$ that are characteristic for spherical membrane segments bound to the particle. For larger values of $z$, the bending energy density drops to zero since the unbound membrane adopts a catenoidal shape with zero mean curvature $M$ and, thus, zero bending energy. Along the contact region at which the membrane detaches from the particle, the interplay of bending and adhesion leads to a small local maximum and a pronounced local minimum in the total energy densities of Fig.~\ref{figure_one_energy-profiles}(c).  

In Fig.~\ref{figure_one}(c), the minimum total energy $E = \int e(z)  \text{d}z$ of the shape profiles with energy density $e(z)$ is shown as a function of the rescaled adhesion energy $u$ for different values of the potential range $\rho$.  The minimum energy $E$ is negative for $\rho>0$ and decreases with $u$. At large values of the rescaled adhesion energy $u$ at which the particles are nearly fully wrapped, the decrease of the minimum energy $E$ is linear in $u$ since the adhesion area and bending energy then are nearly constant. For a fixed value of $u$, the minimum energy $E$ also decreases with increasing potential range $\rho$ because the interplay of bending and adhesion that leads to the local minimum in the energy profiles $e(z)$ of Fig.~\ref{figure_one_energy-profiles}(c) is more pronounced for larger values of $\rho$. In the limit $\rho\to 0$, the total energy $E$ tends towards the black line of Fig.~\ref{figure_one}(c) with $E=0$ for $u<2$ and $E=4\pi \kappa(2  -  u)$ for $u>2$. In this limit, the membrane fully wraps the particle for $u>2$ with bending energy $E_\text{be} = 8\pi\kappa$ and adhesion energy $E_\text{ad}= 4 \pi R^2 U = 4\pi \kappa u$, and the catenoidal membrane neck of zero energy that connects the wrapped membrane segment to the surrounding planar membrane is infinitesimally small.

\thispagestyle{plain}

\section{Cooperative wrapping of particles in long membrane tubes}
\begin{figure}[tp]
\centering
\includegraphics[width=\columnwidth]{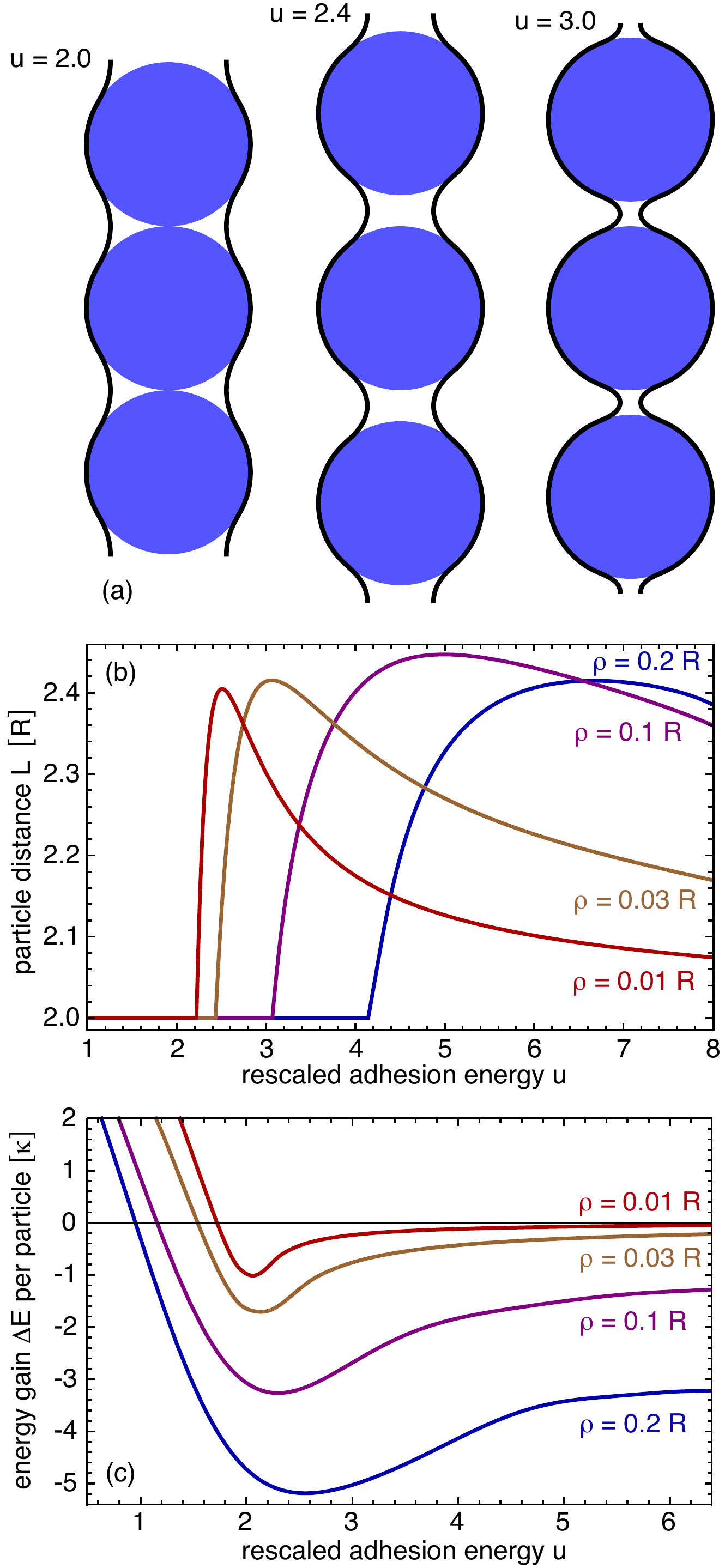}
\caption{(a) Minimum-energy membrane profiles around three central particles of a long tubule for the potential range $\rho = 0.01 R$ and different rescaled adhesion energies $u$. (b) Distance $L\ge 2R$ of neighboring particles in the tube at which the total energy is minimal, and (c) energy gain $\Delta E$ per particle for the cooperative wrapping in long tubes defined in Eq.~(\ref{DeltaE}) as a function of $u$ for various values of the potential range $\rho$. 
}
\label{figure_tube}
\end{figure}

\thispagestyle{plain}

\begin{figure}[tp]
\centering
\includegraphics[width=\columnwidth]{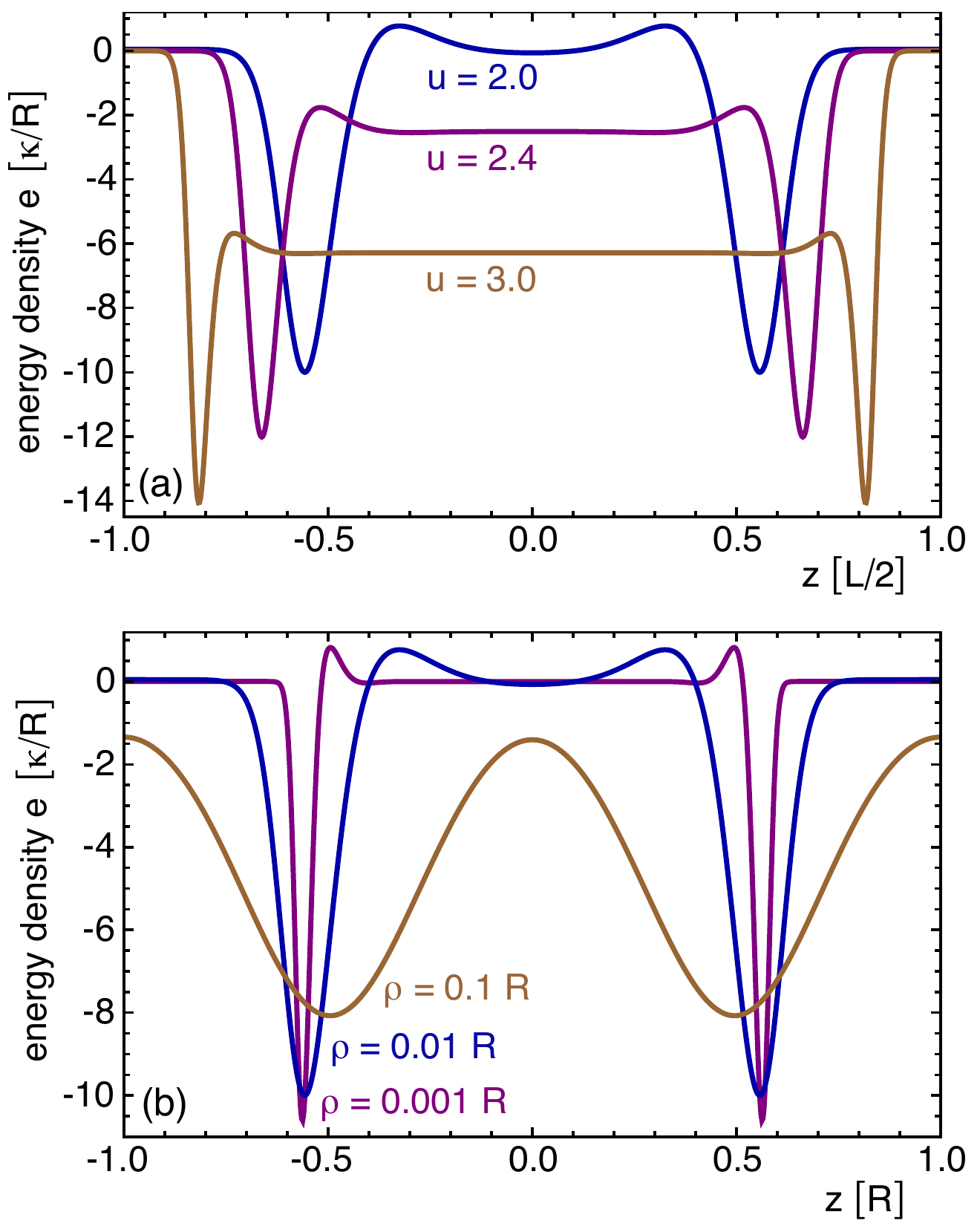}
\caption{Total energy densities $e$ as a function of the coordinate $z$ along the rotational symmetry axis (a) for a single periodic repeat of the three profiles shown in Fig.~\ref{figure_tube}(a) at the potential range $\rho = 0.01 R$, and (b) for single periodic repeats of minimum-energy profiles obtained at the rescaled adhesion energy $u=2.0$ and various values of the potential range $\rho$. The particle center in the single periodic repeat is located at $z = 0$.  
}
\label{figure_tube_energy-profiles}
\end{figure}

In this section, we consider the cooperative wrapping of particles in long tubular membrane structures, with a focus on the membrane energies and shapes around the central particles of such tubes. The membrane energies and shapes around the first and last particles of the tubes will be considered in the next section. Minimum-energy profiles of the membrane around three central particles of long tubes are shown in Fig.~\ref{figure_tube}(a) for the potential range $\rho = 0.01 R$ and three different values of the rescaled adhesion energy $u$. The membrane shapes are periodic along the axis of rotation, and consist of spherical segments bound to the particles that are connected by unbound catenoidal membrane segments between the particles of zero bending energy and, thus, zero total energy (see energy profiles for a single periodic repeat in Fig.~\ref{figure_tube_energy-profiles}(a)). 

\thispagestyle{plain}

The total energy of the shape profiles in Fig.~\ref{figure_tube}(a) is minimized with respect to the distance $L$ of the centers of neighboring particles in the tube. For simplicity, we assume that this distance has to be larger than $2 R$. The particle-particle interaction in the tubes thus is taken to be a hard-sphere interaction with particle radius $R$. At the rescaled adhesion energy $u= 2$ of the left profile in Fig.~\ref{figure_tube}(a)), the total energy $E$ is minimal at the contact distance $L = 2 R$ of the particles. At the values $u= 2.4$ and $u = 3.0$ of the profiles in the center and on the right of Fig.~\ref{figure_tube}(a), the total energy $E$ is minimal at distances larger than $2R$. At these larger values of $u$, the particles are more deeply wrapped by the membrane, and the unbound membrane segments between the particles cannot adopt a catenoidal shape at the contact distance $L = 2 R$, which is energetically unfavorable.
Fig.~\ref{figure_tube}(b) illustrates how the distance $L$ of neighboring particles in the tube depends on $u$ and $\rho$. The particle distance $L$ increases abruptly at a threshold value $u_t$ of the rescaled adhesion energy, and decreases again for large values of $u$. The function $L(u)$ adopts a maximum value at $u =u_m$. This maximum value is slightly larger than $2.4 R$ and depends only weakly on the potential range $\rho$, while the location $u_m$ of the maximum and the threshold value $u_t$ both decrease with $\rho$. 

The energy difference per particle  between the cooperative wrapping in long tubes and the individual wrapping can be defined as
\begin{equation}
\Delta E = E_\text{tube} - E_\text{1p}
\label{DeltaE}
\end{equation}
where $E_\text{tube}$ is the minimum total energy for a central particle in the long tubes considered in this section, and  $E_\text{1p}$ is the minimum total energy of a single wrapped particle shown in Fig.~\ref{figure_one}(c). In Fig.~\ref{figure_tube}(c), the energy difference $\Delta E$ is displayed as a function of the rescaled adhesion energy $u$ for different values of the potential range $\rho$. The energy difference $\Delta E$ is negative for rescaled adhesion energies $u$ larger than a value $u_0$. These negative values of  $\Delta E$ indicate an energy gain for the cooperative wrapping in tubes. The value $u_0$ with $\Delta E = 0$ is located between $u = 1.0$ and $u = 2.0$ and, thus, at values at which single particles are less than half wrapped. For a given potential range $\rho$, the energy difference $\Delta E$ adopts a minimum value at rescaled adhesion energies between $u = 2.0$ and $u = 3.0$. 

\thispagestyle{plain}

A central result is that the energy difference $\Delta E$ between cooperative wrapping and individual wrapping strongly depends on the potential range $\rho$. The minimum values of the energy difference $\Delta E$ per particle are $-5.2\kappa$ for $\rho=0.2R$, $-3.3\kappa$ for $\rho=0.1$,  $-1.7\kappa$ for $\rho=0.03$, and $-1.0\kappa$ for $\rho=0.01 R$. Since typical values of the bending rigidity $\kappa$ range from $10 k_B T$ to $20 k_B T$, these minimum values of  $\Delta E$ are large compared to the thermal energy $k_B T$.  The absolute value of the energy difference $\Delta E$ decreases for intermediate values of the rescaled adhesion energies $u$ between $3.0$ and $6.0$. However, at the large rescaled adhesion energy $u = 6.0$,  the energy differences  $\Delta E= -3.2\kappa$, $-1.3\kappa$, and $-0.24\kappa$ for the potential ranges $\rho = 0.2R$, $0.1R$, and $0.03R$ are still large in magnitude compared to $k_B T$ for typical values of $\kappa$.

\thispagestyle{plain}

The effect of the potential range $\rho$ and rescaled adhesion energy $u$ on the minimum total energies can be understood from the energy densities in the Figs.~\ref{figure_one_energy-profiles} and \ref{figure_tube_energy-profiles}. The minimum total energy $E_\text{tube}$ for a central particle in a long tube is the integral $\int e(z)\text{d}z$ of the total energy densities in Fig.~\ref{figure_tube_energy-profiles}, and the minimum total energy $E_\text{1p}$ for a single wrapped particle is the integral of the total energy densities of Fig.~\ref{figure_one_energy-profiles}(c). Bound, spherical membrane segments that are located in the minimum of the adhesion potential have the constant, $z$-independent energy density $e(z) = 2\pi (2- u) \kappa/R= e_\text{sphere}$. The total energy densities of a tube particle in Fig.~\ref{figure_tube_energy-profiles}(a) adopt the value $e(z)= e_\text{sphere}$ around $z= 0$, which corresponds to the center of the particle. The total energy densities of Fig.~\ref{figure_one_energy-profiles}(c) for a single particle with center located at $z = R$ adopt the value $e(z)= e_\text{sphere}$   for small values of $z$. Unbound, catenoidal membrane segments have the energy density zero.  Such catenoidal segments are located around the values $z = -L/2$ and $z = L/2$ of Fig.~\ref{figure_tube_energy-profiles}(a) in between the tube particles, and at large values of $z$ in Fig.~\ref{figure_one_energy-profiles}(c) for a single wrapped particle. Along the contact regions at which the bound membrane detaches from the particle, the interplay of bending and adhesion energies leads to local minima in the total energy profiles. 

\thispagestyle{plain}

At the rescaled adhesion energy $u = 2.0$,  the energy difference $\Delta E$ between the cooperative wrapping in tubes and the individual wrapping of the particles results from the contact regions at which the membrane detaches from the particles, because the energy densities are both zero for bound and unbound membrane segments at this value of $u$.  Since each particle in a tube has two such contact regions, the profiles in Fig.~\ref{figure_tube_energy-profiles} exhibit two minima, while the profiles of the single particles in \ref{figure_one_energy-profiles}(c) with a single contact region just exhibit one minimum. With increasing potential range $\rho$, the minima in the  energy profiles become broader (see Fig.~\ref{figure_tube_energy-profiles}(b)). For the potential range  $\rho = 0.1 R$, the interplay between bending and adhesion affects the whole profile $e(z)$ at the rescaled adhesion energy $u = 2.0$ (see brown profile in Fig.~\ref{figure_tube_energy-profiles}(b)).

\begin{figure}[tp]
\centering
\includegraphics[width=\columnwidth]{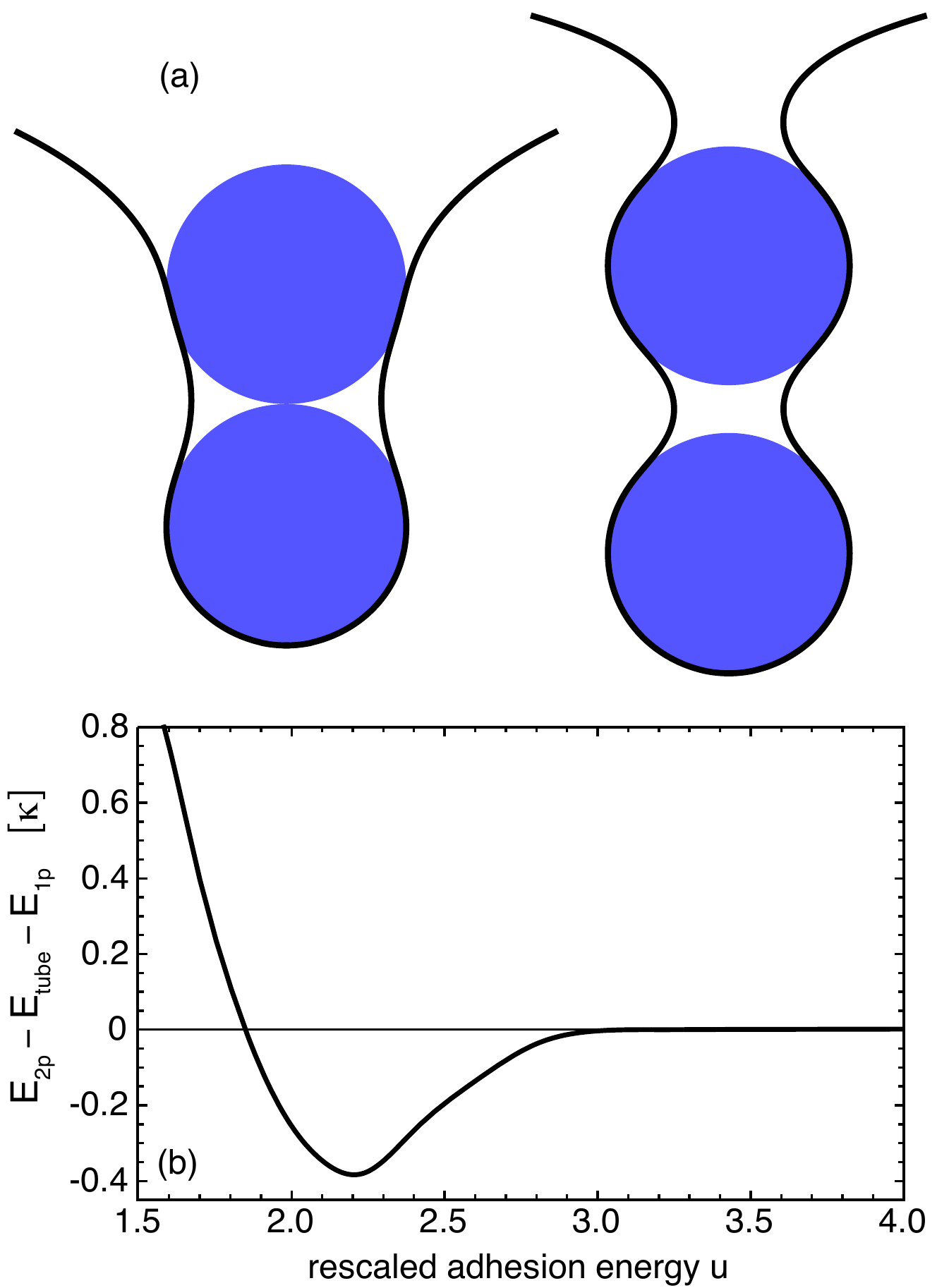}
\caption{(a) Minimum-energy profiles of membrane tubes with two particles at the potential range $\rho = 0.1 R$ and the rescaled adhesion energies $u = 2.0$ (left) and $u = 4.0$ (right). (b) Energy difference $E_\text{2p} - E_\text{1p} - E_\text{tube}$ between the minimum total energy $E_\text{2p}$ of a two-particle tube and the sum $E_\text{1p} + E_\text{tube}$ of the minimum total energies for a single wrapped particle and a central particle in a long tube as a function of $u$ for $\rho = 0.1 R$.
}
\label{figure_2p}
\end{figure}

\thispagestyle{plain}

\section{Cooperative wrapping in tubes of two or more particles }

The energy contribution of the first and last particles in a tube can be understood from the shapes and minimum energies of tubular structures with two or more particles. Fig.~\ref{figure_2p}(a) displays the minimum-energy profiles of a two-particle tube at the potential range $\rho = 0.1 R$. At the rescaled adhesion energy $u=2.0$, the two particles have the contact distance $L = 2 R$ in their minimum-energy configuration (left profile). At $u = 4.0$, the total energy is minimal for the particle distance $L = 2.4 R$ (right profile). At the potential range $\rho = 0.1 R$, the minimal total energy $E_\text{2p}$ of a two-particle tube can be approximated as  
\begin{equation}
E_\text{2p} \simeq  E_\text{1p} + E_\text{tube} 
\label{E2p}
\end{equation}
for rescaled adhesion energies $u\gtrsim 3$ where $E_\text{1p}$ is the minimal total energy for a single particle, and $E_\text{tube}$ is the minimum total energy for a particle in a long tube (see Fig.~\ref{figure_2p}(b)). Eq.~(\ref{E2p}) holds because the membrane profiles around the two inner half spheres of the particles that face each other is similar to the profile of central particles in a long tube, while the profile segments around the two outer half spheres that face away from each other are similar to the profile of a single wrapped particle for intermediate and large rescaled adhesion energies (see right profile of Fig.~\ref{figure_2p}(a)). An extension of this argument to tubes of more than two particles leads to the minimum total energy 
\begin{equation}
E_\text{np} \simeq E_\text{1p} + (n-1)E_\text{tube}
\end{equation}
of a tubular protrusion with $n$ particles for intermediate and large rescaled adhesion energies. The energy difference per particle between the cooperative wrapping in an $n$-particle membrane tube compared to individual wrapping thus is $(n-1)\Delta E/n$ where $\Delta E$ is the energy difference for a central particle of a long tube shown in Fig.~\ref{figure_tube}(c).

\thispagestyle{plain}

\section{Discussion and Conclusions}

In this article, we have determined the energy gain $\Delta E$ for the cooperative wrapping of particles by membrane tubes. We have found that this energy gain strongly depends on the ratio $\rho/R$ of the potential range $\rho$ and particle radius $R$ because of a favorable interplay between bending and adhesion that becomes more pronounced with increasing $\rho/R$ (see Fig.~6(b)). This interplay mainly occurs in the contact regions in which the membrane detaches from the particles, in particular for larger values of the rescaled adhesion energy $u$  (see Fig.~6(a)). The cooperative wrapping in tubes then is favorable because a particle in a tube has two such contact regions with the membrane, while a single wrapped particle only has one contact region. 

We have considered here the wrapping of nanoparticles by large planar membranes with negligible tension $\sigma$. In general, the membrane tension is negligible if the crossover length $\sqrt{\kappa/\sigma}$ is large compared to the particle radius $R$, because the elastic energy of the membranes then is dominated by the bending energy \cite{Lipowsky95}. On length scales larger then the crossover length $\sqrt{\kappa/\sigma}$, the elastic energy is dominated by the tension. For typical values of the bending rigidity $\kappa$ between 10 and 20 $k_B T$ \cite{Seifert95} and a membrane tension  $\sigma$ of a few $\mu\text{N}/\text{m}$ \cite{Simson98}, for example, the crossover length adopts values between 100 and 200 nm.

In experiments, the aggregation of nanoparticles in solution is typically prevented by repulsive interactions between the nanoparticles, e.g.~by electrostatic interactions if the particles are charged. In general, such repulsive interactions can affect the energies of the particle-filled membrane tubes, in particular if neighboring particles in these tubes are in contact. For simplicity, we have considered here nanoparticles that exhibit only repulsive hard-sphere interactions.  
 However, it is important to note that the neighboring particles in our tubes are not in contact at intermediate and large values of the rescaled adhesion energy $u$ (see Figs.~5(a) and (b)). For such rescaled adhesion energies, repulsive interactions between the particles only affect our results if their interaction range is larger than the distance between the surfaces of neighboring particles in the tubes. 
  
In general, the energy gain $\Delta E$ for the cooperative wrapping of particles by membrane tubes also depends on the particle shape. For prolate particles, for example, we expect larger absolute values of $\Delta E$ than for the spherical particles considered here because the more strongly curved tips of prolate particles do not have to be wrapped in membrane tubes, which provides an additional advantage compared to the individual wrapping of the particles. If prolate particles are wrapped individually, one of the tips is enclosed by the membrane in deeply wrapped states of the particles \cite{Bahrami13}. For oblate particles, we expect tubular structures in which the more strongly curved edges of neighboring particles face each other, because then at least parts of these edges do not have to be wrapped. Such tubular structures of oblate particles do not exhibit rotational symmetry.

\thispagestyle{plain}

The large energy gain for the cooperative wrapping of particles implies strongly attractive elastic interactions that are mediated by the membrane. These elastic interactions result from the fact that the minimum total energy of two or more adhering particles depends on the particle distances. At the optimal distance $L$ for the cooperative wrapping of the particles by membrane tubes, the total energy is significantly lower than at large distances at which the particles are wrapped individually by the membrane (see Fig.~\ref{figure_tube}). Membrane shape fluctuations can induce additional attractive interactions between adsorbed particles since the particles suppress such fluctuations in their adhesion zones. However, these fluctuation-induced, entropic interactions are of the order of the thermal energy $k_B T$ \cite{Goulian93,Golestanian96,Weikl01b,Lin11} and thus significantly weaker than the elastic energy gain $\Delta E$ for the cooperative wrapping displayed in Fig.~\ref{figure_tube}(c), since the bending rigidity $\kappa$ of the membranes is of the order of 10 $k_B T$ \cite{Seifert95}. In addition to the weak entropic interactions, the suppression of membrane shape fluctuations in the adhesion zone of the particles effectively reduces the adhesion energy $U$ per area \cite{Lipowsky86,Helfrich78}.

\thispagestyle{plain}

\footnotesize
\providecommand*{\mcitethebibliography}{\thebibliography}
\csname @ifundefined\endcsname{endmcitethebibliography}
{\let\endmcitethebibliography\endthebibliography}{}

\thispagestyle{plain}


\end{document}